\documentclass[preprint,authoryear,1p,12pt]{elsarticle}

\usepackage{graphicx}

\usepackage{amssymb}
\usepackage{amsthm}
\usepackage{amsmath}

\usepackage{algorithm}
\usepackage{algpseudocode}

\usepackage{multirow}
\usepackage{array}

\journal{Applied Mathematics and Computation}

\begin{document}

\begin{frontmatter}



\title{MOCSA: multiobjective optimization by conformational space annealing}

\address[kias]{School of Computational Sciences, Korea Institute for Advanced Study, Seoul, Korea}

\author[kias]{Sangjin Sim, Juyong Lee and Jooyoung Lee}


\begin{abstract}
  We introduce a novel multiobjective optimization algorithm based on the conformational space annealing (CSA) algorithm, MOCSA. It has three characteristic features: (a) Dominance relationship and distance between solutions in the objective space are used as the fitness measure, (b) update rules are based on the fitness as well as the distance between solutions in the decision space and (c) it uses a constrained local minimizer. We have tested MOCSA on 12 test problems, consisting of ZDT and DTLZ test suites. Benchmark results show that solutions obtained by MOCSA are closer to the Pareto front and covers a wider range of the objective space than those by the elitist non-dominated sorting genetic system (NSGA2).
\end{abstract}


\end{frontmatter}


\section{Introduction}
\label{sec:intro}
The multiobjective optimization problem (MOOP) is to optimize two or more objective functions simultaneously, subject to given constraints.
The multiobjective optimization can be applied to problems where the final decision should be made considering two or more conflicting objectives.
MOOP occurs in various fields such as industrial design, finance, management and many engineering areas. 
Practical goals in these fields can be generalized in such a way that the cost of a process is minimized while the quality of its product is maximized.
The primary goal is to find a set of solutions that any individual objective function cannot be improved without deteriorating the other objective functions, and such a set is called a Pareto set.
For efficient decision making, a set of generated solutions ($\mathcal{GS}$) should meet two conditions: It should be as close to the Pareto front as possible and the solutions should be distributed as widely as possible. 

Evolutionary algorithm (EA) is one of the most popular and successful approaches to solve MOOPs~\citep{deb2001multi,coello2007evolutionary}. A number of EA-based algorithms have been suggested including the vector evaluated genetic algorithm (VEGA)~\citep{schaffer1985multiple}, the niched Pareto genetic algorithm (NPGA)~\citep{horn1994niched}, the nondominated sorting genetic algorithm \MakeUppercase{\romannumeral 2} (NSGA2)~\citep{deb2000fast}, the strength Pareto evolutionary algorithm \MakeUppercase{\romannumeral 2} (SPEA2)~\citep{zitzler2001spea2}, the mimetic Pareto archived evolution strategy (M-PAES)~\citep{knowles2000m} and micro genetic algorithm (micro-GA)~\citep{coello2001micro}. Among them, NSGA2 and SPEA2 are arguably the most widely used methods. Other approaches include simulated annealing (SA)~\citep{suman2005survey,nam2000multiobjective}, tabu search~\citep{hansen1997tabu,gandibleux1997tabu}, particle swarm optimization (PSO)~\citep{parsopoulos2002particle,coello2002mopso}, immune algorithm (IA)~\citep{coello2002approach,gao2011hybrid,luh2003moia}, ant system~\citep{doerner2004pareto,baran2003multiobjective} and cultural algorithm~\citep{coello2003evolutionary}.

Conformational Space Annealing (CSA) is a highly efficient single-objective global optimization algorithm which incorporates advantages of genetic algorithm and SA. It has been successfully applied to diverse single-objective optimization problems in physics and biology, such as protein structure modeling~\citep{Lee1999a,Pillardy2001,Liwo1999,joo2009all}, finding the minimum energy solution of a Lenard-Jones cluster~\citep{Lee2003}, multiple sequence alignment~\citep{joo2008multiple} and the community detection problem~\citep{Lee2012} on networks. In these studies, CSA is shown to perform more efficient sampling using less computational resources than the conventional Monte-Carlo (MC) and SA methods. 

Here, we introduce a new multiobjective optimization algorithm by using CSA, MOCSA. Compared to existing EAs, MOCSA has the following distinct features: (a) The ranking system considers the dominance relationship and the distance between solutions in the objective space, (b) solutions are updated by using a dynamically varying distance cutoff measure to control the diversity of the sampling in the decision space, and (c) a gradient-based constrained minimizer is utilized for local search.

The remainder of this paper is organized as follows. In section 2, the definition of MOOP and related terms are described. In section 3, details of MOCSA is presented. Numerical results and the comparison between MOCSA and NSGA2 on various test problems are presented in section 4. The final section contains the conclusion.

\section{Problem statement}
\label{problem}

The mathematical definition of a MOOP can be defined as follows, 
\begin{align*}
  \min_{\mathbf{u}}\mathbf{v} &= \min_{\mathbf{u}}\mathbf{f}(\mathbf{u}) = \min_{\mathbf{u}}\left(f_{1}(\mathbf{u}), f_{2}(\mathbf{u}), \dotsc, f_{m}(\mathbf{u})\right), \\
  \text{s.t. }\mathbf{u}&=(u_{1}, u_{2}, \dotsc, u_{n}) \in U \subset R^{n} \\
  \mathbf{v}&=(v_{1}, v_{2}, \dotsc, v_{m}) \in V \subset R^{m}
\end{align*}
where $\mathbf{u}$ is the decision vector, $U$ the decision space, $\mathbf{v}$ the objective vector and $V$ the objective space. 
Due to the presence of multiple objective functions, a final solution of MOOP consists of a set of non-dominated solutions instead of a single point. 
The notion of \emph{dominance} and related terms are defined below.

\newdefinition{paretodominance}{Definition}
\begin{paretodominance}
  A decision vector $\mathbf{u_{1}}$ is said to dominate another solution $\mathbf{u_{2}}$ (denoted by $\mathbf{u_{1}} \prec \mathbf{u_{2}}$), if and only if 
  \begin{equation}
    \forall i \in \{1, \dotsc, m\}:f_{i}(\mathbf{u_{1}}) \le f_{i}(\mathbf{u_{2}}) \land \exists k \in \{1, \dotsc, m\}:f_{k}(\mathbf{u_{1}}) < f_{k}(\mathbf{u_{2}}).
  \end{equation}
\end{paretodominance}

\newdefinition{paretooptimal}[paretodominance]{Definition}
\begin{paretooptimal}
  A solution $\mathbf{u}$ is said to be non-dominated by any other solutions (a Pareto optimal solution) if and only if 
  \begin{equation}
    \neg\exists\mathbf{u^{*}} \in U: \mathbf{u^{*}} \prec \mathbf{u}.
  \end{equation}
\end{paretooptimal}

\newdefinition{paretooptimalset}[paretodominance]{Definition}
\begin{paretooptimalset}
  For a given MOOP, a Pareto optimal set in the decision space, $\mathcal{PS}$, is defined as
  \begin{equation}
    \mathcal{PS}:=\{\mathbf{u}|\neg\exists\mathbf{u^{*}} \in U: \mathbf{u^{*}} \prec \mathbf{u}\}.
  \end{equation}
\end{paretooptimalset}

\newdefinition{paretooptimalfront}[paretodominance]{Definition}
\begin{paretooptimalfront}
  For a given MOOP, a Pareto optimal set in the objective space, $\mathcal{PF}$, is defined as
  \begin{equation}
    \mathcal{PF}:=\{\mathbf{f}(\mathbf{u})|\mathbf{u}\in\mathcal{PS}\}.
  \end{equation}
\end{paretooptimalfront}

Since the size of Pareto optimal front, $\mathcal{PF}$ is infinite in general, which is impossible to obtain in practice, practical algorithms for MOOP yield a set of non-dominated solutions of a finite size. It should be noted that $\mathcal{PF}$ is always a non-dominated set by definition while a non-dominated set of solutions generated by an algorithm, which is denoted as a $\mathcal{GS}$, may not be a subset of $\mathcal{PF}$. 

\section{Description of conformational space annealing}
\label{sec:CSA}
{
  Here, a new multiobjective optimization algorithm based on CSA is described. The CSA was initially developed to obtain the protein structure with the minimum potential energy, \emph{i.e.}, to solve a single objective optimization problem. CSA has been successfully applied to various kinds of optimization problems with modification. The general framework of CSA is shown in Figure~\ref{CSA_flow_chart}, and the description of MOCSA is given in Algorithm~\ref{CSA}.

\begin{figure}[!ht]
  \begin{center}
    \includegraphics[width=5.5in]{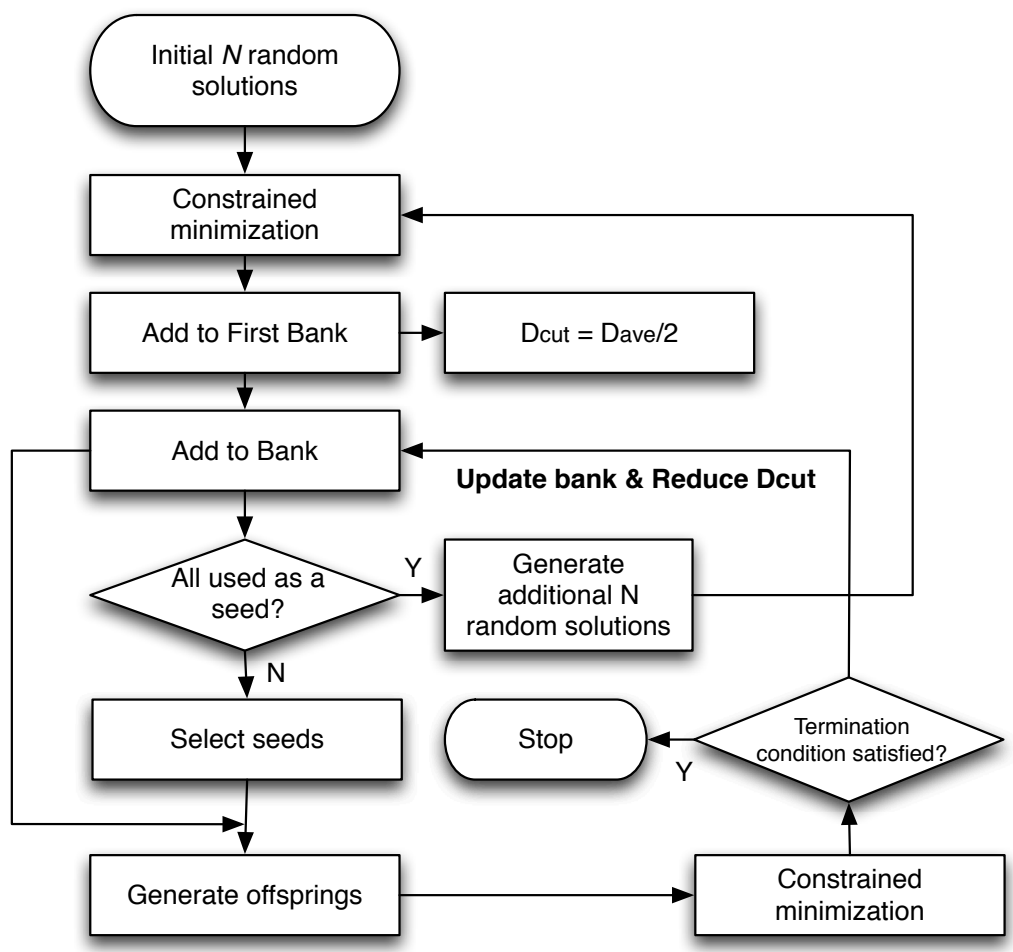}
  \end{center}
  \caption{
    {\bf Flow chart of CSA is shown.} 
  }
  \label{CSA_flow_chart}
\end{figure}

\begin{algorithm}
  \caption{Multiobjective CSA algorithm}
  \label{CSA}
  \begin{algorithmic}[1]

    \Procedure{MOCSA}{$N, N_{s}, N_{c}, N_{m}, G = 300, \alpha$}
    \State Initialize the bank, $\mathbb{P}$, with $N$ random individuals
    \State Minimize($\mathbb{P}$) using a constrained local minimizer
    \State Initialize seed flags of all individuals to zeros : $s(i) \gets 0,  i=1, \dotsc, N$
    \State Get average distance, $D_{ave}$, between all pairs of individuals and set $D_{cut}$ as $D_{ave}/2$: $D_{cut} \gets D_{ave}/2$
    \State Initialize generation counter to zero : $g \gets 0$
    \State Initialize the reserve bank, $\mathbb{R}$, to an empty set

    \While {$g < G$}
    \If{$s(i) == 1 \quad \forall i$}
    \State Generate $N$ random individuals, $\mathbb{P'}$
    \State Minimize($\mathbb{P'}$)
    \State $\mathbb{P} \gets \mathbb{P} \cap \mathbb{P'}$\Comment Expand search space
    \EndIf
    \State Evaluate Fitness of $\mathbb{P}$
    \State Select $N_{s}$ seeds among individuals with $s(i)=0$ and set $s(i)$ to 1
    \State $\mathbb{T}_{c} \gets $ Generate $N_{s}N_{c}$ trial solutions by crossover
    \State $\mathbb{T}_{m} \gets $ Generate $N_{s}N_{m}$ trial solutions by mutation
    \State $\mathbb{T} \gets \mathbb{T}_{c} \cup \mathbb{T}_{m}$\Comment Trial solutions
    \If{$G \% 5 == 0$}
      \State Minimize($\mathbb{T}$)
    \EndIf
    \State Update($\mathbb{P}, \mathbb{T}, \mathbb{R}$)
    \State $D_{cut} \gets max(\alpha D_{cut}, D_{ave}/5)$  \Comment Reduce $D_{cut}$ 
    \State $g \gets g+1$
    \EndWhile\label{CSAendwhile}
    \EndProcedure

  \end{algorithmic}
\end{algorithm}

\subsection{Conformational space annealing}
{
\indent CSA is a global optimization method which combines essential ingredients of three methods: Monte Carlo with minimization (MCM)~\citep{Li1987monte},
 genetic algorithm (GA)~\citep{goldberg1989genetic}, and SA~\citep{Kirkpatrick1983optimization}. 
 As in MCM, we consider only the solution/conformational space of local minima; in general, all solutions are minimized by a local minimizer. 
 As in GA, we use a set of $N$ solutions (called \emph{bank} in CSA, denoted as $\mathbb{P}$) collectively, 
 and we generate offsprings from the bank solutions by cross-over and mutation. 
 Finally, as in SA, we introduce a distance parameter $D_{cut}$, which plays the role of the temperature in SA. 
 In CSA, each solution is assumed to represent a hyper-sphere of radius $D$ in the decision space. 
 Diversity of sampling is directly controlled by introducing a distance measure between two solutions and comparing it with $D_{cut}$, to prevent two solutions from approaching 
 too close to each other in the decision space. 
 Similar to the manipulation of temperature in SA, the value of $D_{cut}$ is initially set to a large value and is slowly reduced to a smaller value in CSA; hence the name conformational space annealing.
 \newline
 \indent Compared to the conventional EA for multiobjective problems, MOCSA has three distinct features;
(a) a ranking algorithm which considers the dominance relationship as well as the distance between solutions in the objective space, 
(b) an update rule with a dynamically varying distance cutoff measure to control the size of search space and to keep the diversity of sampling in the decision space
and (c) the usage of a gradient-based constrained minimizer, feasible sequential quadratic programming (FSQP), for local search.
 \newline
 \indent In CSA, we first initialize the \emph{bank}, $\mathbb{P}$, with $N=50$ random solutions which are subsequently minimized by FSQP constrained minimizer.
 The solutions in the bank are updated using subsequent solutions found during the course of optimization. 
 The initial value of $D_{cut}$ is set as $D_{avg}/2$, where $D_{avg}$ is the average distance in the \emph{decision} space between two solutions at the initial stage.
 A number of solutions (20 in this study) in the bank are selected as \emph{seeds}.
 For each seed, 30 trial solutions are generated by cross-over between the seed and randomly chosen solutions from the bank. 
 Additional 5 are generated by mutation of the seed. 
 It should be noted that if a solution is used as a seed and not replaced by a offspring, it is excluded from the subsequent seed selection.
 The generated offsprings are locally minimized by FSQP which guarantees to improve a subset of objective functions 
 without deteriorating the others and without violating given constraints.
 To limit the computational usage, the minimization is performed only once per every five generation steps (see Algorithm~\ref{CSA}).
 
 \indent Offsprings are used to update the bank, and detailed description on the updating rule is provided in Section~\ref{sec:update}. 
 Once all solutions in the bank are used as seeds without generating better solutions,
 implying that the procedure might have reached a deadlock, we reset all bank solutions to be eligible for seeds again and repeat another round of the search procedure. 
 After this additional search reaches a deadlock again, we expand our search space by adding additional 50 randomly generated
 and minimized solutions to the bank ($N=N+50$), and repeat the whole procedure until a termination criterion is satisfied. The maximum number of generation is set to $G$.
 Typically, with $G=300$ in this study, MOCSA is terminated before a deadlock occurs with the final bank size of $N=50$.
}

\subsection{Fitness function}
\label{sec:fitness}
{
  For a given set of generated solutions, $\mathcal{GS}$, the fitness of solution $i$ is evaluated in terms of $n_i$, $m_i$ and $d^{12}_{i}$. $n_i$ is the number of solutions in $\mathcal{GS}$ which dominate $i$. $m_i$ is the number of solutions in $\mathcal{GS}$ dominated by $i$. $d^{12}_{i}$ is the sum of distances from $i$ to its nearest and second nearest neighbors in $\mathcal{GS}$ in the objective space. The relative fitness between two solutions, $i$ and $j$, is determined by the comparing function shown in Algorithm~\ref{Fitness}. With a set of non-dominated solutions all values of $n_i$ and $m_i$ become zeros and the solution with the least value of $d^{12}$ is considered as the worst.
}

\subsection{Update rule}
\label{sec:update}
{
  The solutions generated by crossover and mutation are locally minimized by FSQP constrained minimizer and we call them trial solutions.
  Each trial solution $\mathbf{t}$, is compared with the bank $\mathbb{P}$ for update procedure as shown in Algorithm~\ref{Update}.
  First, $\mathbf{w}$, the closest solution in $\mathbb{P}$ from $\mathbf{t}$ in the \emph{decision} space is identified.
  If there exist dominated solutions in $\mathbb{P}$, the closest conformation search is performed only among them. 
  Otherwise, $\mathbb{P}$ is a set of non-dominated solutions, and all in $\mathbb{P}$ are considered. 
  Once $\mathbf{w}$ is found, the distance $D$ in the decision space between $\mathbf{t}$ and $\mathbf{w}$ is calculated. 
  If $D > D_{cut}$, the current cutoff distance, which indicates that $\mathbf{t}$ lies in a newly sampled region in the decision space, 
  remote from the existing solutions in $\mathbb{P}$, the dominance relationship between $\mathbf{t}$ and the worst solution in $\mathbb{P}$, $\mathbf{u}$, is compared. 
  If $D \leq D_{cut}$, $\mathbf{t}$ is compared with $\mathbf{w}$.
  The selection procedure, described in Section~\ref{sec:select}, is performed to determine which solution should be kept in $\mathbb{P}$.
  At each iteration step, $D_{cut}$ is reduced with a pre-determined ratio, $\alpha$. 
  After $D_{cut}$ reaches to its final value, $D_{avg}/5$, it is kept constant.
}

\subsection{Selection procedure}
\label{sec:select}
{
  In Algorithm~\ref{Update}, for a given trial solution $\mathbf{t}$ and a solution $\mathbf{x}$ in $\mathbb{P}$, $\mathbb{P}$ is updated as follows.
  If $\mathbf{t}$ dominates $\mathbf{x}$, $\mathbf{t}$ replaces $\mathbf{x}$. If $\mathbf{t}$ is dominated by $\mathbf{x}$, $\mathbf{t}$ is discarded and $\mathbf{x}$ stays in $\mathbb{P}$. If there is no dominance relationship between $\mathbf{t}$ and $\mathbf{x}$ and if $\mathbf{x}$ is better than $\mathbf{t}$ by Algorithm~\ref{Fitness}, $\mathbf{t}$ is discarded and $\mathbf{x}$ stays in $\mathbb{P}$. 
  Finally, when $\mathbf{t}$ is better than $\mathbf{x}$ without dominance relationship between them, Algorithm~\ref{Select} is used.

  For the selection procedure, we introduce an additional set of non-dominated solutions, the reserve bank, $\mathbb{R}$.
  Due to the limited size of the bank, we may encounter a situation where a solution exists in $\mathbb{P}$ which is not dominated by the current bank, 
  but dominated by a solution eliminated from the bank in an earlier generation. 
  To solve this problem, non-dominated solutions eliminated from $\mathbb{P}$ are stored up to 500 in $\mathbb{R}$, which is conceptually similar to an \emph{archive} in other EAs~\citep{knowles2000m,zitzler2001spea2}. The difference is that $\mathbb{R}$ is used only when more than half of the solutions in $\mathbb{P}$ are non-dominated solutions because CSA focuses more on diverse sampling rather than optimization at the early stage of the optimization. Note that $\mathbb{R}$ keeps only non-dominated solutions.
}

\begin{algorithm}
\caption{Comparing function}
\label{Fitness}
\begin{algorithmic}[1]
  \Function{Compare}{$i, j$} 
  \State $n_{i (j)}\gets $ Number of solutions in $\mathbb{P}$ which dominate $\mathbf{i}$($\mathbf{j}$)         
  \State $m_{i (j)}\gets $ Number of solutions in $\mathbb{P}$ which are dominated by $\mathbf{i}$($\mathbf{j}$) 
  \State $d^{12}_{i (j)}\gets $ The sum of distances from $i$ to the nearest and second nearest neighbors in $\mathbb{P}$ in the \emph{objective} space

  \If {$n_i < n_j$}
  \State \Return $\mathbf{i}$ is better
  \ElsIf {$n_i > n_j$}
  \State \Return $\mathbf{j}$ is better
  \Else\Comment $n_i == n_j$

  \If {$m_i > m_j$}
  \State \Return $\mathbf{i}$ is better
  \ElsIf {$m_i < m_j$}
  \State \Return $\mathbf{j}$ is better
  \Else\Comment $m_i == m_j$
  
  \If {$d^{12}_{i} > d^{12}_{j}$}
  \State \Return $\mathbf{i}$ is better
  \Else
  \State \Return $\mathbf{j}$ is better
  \EndIf
  
  \EndIf
  \EndIf
  \EndFunction
\end{algorithmic}
\end{algorithm}
}

\begin{algorithm}
\caption{Update procedure}
\label{Update}
\begin{algorithmic}[1]

  \Procedure{Update}{$\mathbb{P}, \mathbb{T}, \mathbb{R}$}

  \For {$\mathbf{t}$ in $\mathbb{T}$}
  \If {all solutions in $\mathbb{P}$ are non-dominated}
  \State $\mathbf{w} \gets$ Nearest solution to $\mathbf{t}$ in $\mathbb{P}$ in the \emph{decision} space
  \Else
  \State $\mathbf{w} \gets$ Nearest solution to $\mathbf{t}$ among \emph{dominated} solutions in $\mathbb{P}$ in the \emph{decision} space
  \EndIf

  \If {$distance(\mathbf{t},\mathbf{w}) > D_{cut}$} \Comment distance in the decision space
  \State $\mathbf{u} \gets$ Worst solution in $\mathbb{P}$
  \State $\mathbf{x} \gets \mathbf{u}$
  \Else
  \State $\mathbf{x} \gets \mathbf{w}$
  \EndIf
  
  \If {$\mathbf{t} \prec \mathbf{x}$}     \Comment $\mathbf{t}$ dominates $\mathbf{x}$
  \State $\mathbf{t}$ replaces $\mathbf{x}$
  \ElsIf {$\mathbf{t} \succ \mathbf{x}$}  \Comment $\mathbf{x}$ dominates $\mathbf{t}$
  \State $\mathbf{x}$ stays in $\mathbb{P}$ and $\mathbf{t}$ is discarded
  \ElsIf {$\mathbf{x}$ is better than $\mathbf{t}$ by Algorithm~\ref{Fitness}} 
  \State $\mathbf{x}$ stays in $\mathbb{P}$ and $\mathbf{t}$ is discarded
  \Else
  \State SELECT($\mathbf{t}, \mathbf{x}$) 
  \EndIf

  \EndFor
  \EndProcedure
\end{algorithmic}
\end{algorithm}

\begin{algorithm}
\caption{Select procedure}
\label{Select}
\begin{algorithmic}[1]

  \Procedure{Select}{$\mathbf{t}, \mathbf{x}$}\Comment $\mathbf{t} \in \mathbb{T}$, $\mathbf{x} \in \mathbb{P}$, $\mathbf{t}$ is better than $\mathbf{x}$
  
  \If {$|\mathcal{ND(\mathbb{P})}| < |\mathbb{P}|/2$ or $\mathbf{x}$ is dominated by $\mathbf{x'} \in \mathbb{P}$}
  \State  $\mathbf{t}$ replaces $\mathbf{x}$ in $\mathbb{P}$
  \Else \Comment $\mathbf{x}$ is non-dominated in $\mathbb{P}$ and $\mathbb{R}$ is used

  \If {$\mathbf{t}$ is not dominated by $\mathbb{R}$}
  \State  $\mathbf{t}$ replaces $\mathbf{x}$ in $\mathbb{P}$
  \Else \Comment $\mathbf{t}$ is dominated by $\mathbb{R}$
  \State $\mathbf{u} \gets$ Nearest dominating solution to $\mathbf{t}$ in $\mathbb{R}$ in the \emph{objective} space
  \State Move $\mathbf{u}$ from $\mathbb{R}$ to $\mathbb{P}$
  \EndIf

  \State Move $\mathbf{x}$ from $\mathbb{P}$ to $\mathbb{R}$
  \State $\mathbb{R} \gets \mathcal{ND(\mathbb{R})}$
  \EndIf

  \EndProcedure
\end{algorithmic}
\end{algorithm}


\newsavebox\zdtone
\begin{lrbox}{\zdtone}
  \begin{minipage}{0.2\textwidth}
    \begin{align*}
      f_1 &= y_1 \\
      g   &= 1 + 9\dot\sum_{i=1}^{k}{z_{i}/k} \\
      h(f_1,g)&=1-\sqrt{f_{1}/g} \\
    \end{align*} 
  \end{minipage}
\end{lrbox}

\newsavebox\zdtfourdomain
\begin{lrbox}{\zdtfourdomain}
  \begin{minipage}{0.2\textwidth}
    \begin{align*}
      &y_1  \in [0,1] \\
      &z_1, \dotsc, z_k  \in [-5,5]
    \end{align*}
  \end{minipage}
\end{lrbox}

\newsavebox\zdtsix
\begin{lrbox}{\zdtsix}
  \begin{minipage}{0.2\textwidth}
    \begin{align*}
      f_1 & = 1 - exp(-4y_1)sin^6(6\pi y_1)\\
      g   & = 1+9\dot(\sum^{k}_{i=1}{z_{i}/k)})^{0.25} \\
      h   & = 1 - (f_{1}/g)^{2} \\
    \end{align*} 
  \end{minipage}
\end{lrbox}

\newsavebox\dtlzone
\begin{lrbox}{\dtlzone}
  \begin{minipage}{0.2\textwidth}
    \begin{align*}
      f_1 & = \frac{1}{2}(1+g)\prod^{j-1}_{i=1} y_i \\
      f_{m=2:j-1} & = \frac{1}{2}(1+g)\left(\prod^{j-m}_{i=1} y_i \right)(1-y_{j-m+1}) \\
      f_j & = \frac{1}{2}(1+g)(1-y_1) \\
      g & = 100[k+\sum^{k}_{i=1}\left((z_i-0.5)^2 - cos(20\pi (z_i-0.5))\right)] \\
    \end{align*} 
  \end{minipage}
\end{lrbox}

\newsavebox\dtlztwo
\begin{lrbox}{\dtlztwo}
  \begin{minipage}{0.2\textwidth}
    \begin{align*}
      f_1 & = (1+g)\prod^{j-1}_{i=1} cos(\pi y_i/2) \\
      f_{m=2:j-1} & = (1+g)\left(\prod^{j-m}_{i=1} cos(\pi y_i/2) \right)sin(\pi y_{j-m+1}/2) \\
      f_j & = (1+g)sin(\pi y_1/2) \\
      g & = \sum_{i=1}^{k}(z_i - 0.5)^2 \\
    \end{align*} 
  \end{minipage}
\end{lrbox}

\newsavebox\dtlzseven
\begin{lrbox}{\dtlzseven}
  \begin{minipage}{0.2\textwidth}
    \begin{align*}
      f_{m=1:j-1} & = y_{m} \\
      f_j & = (1+g)\left(j-\sum_{i=1}^{j-1}[\frac{f_i}{1+g}(1+sin(3\pi f_i))]\right) \\
      g & = 1+9\sum_{i=1}^{k}z_i/k  \\
    \end{align*} 
  \end{minipage}
\end{lrbox}

\newsavebox\wfgslinear
\begin{lrbox}{\wfgslinear}
  \begin{minipage}{0.2\textwidth}
    \begin{align*}
      \text{linear}_{1}(y_1,\dotsc,y_{j-1})=\prod^{j-1}_{i=1}y_i\\
    \end{align*} 
  \end{minipage}
\end{lrbox}
    
\section{Test suites}
\label{sec:testproblem}

{
  For the benchmark test of MOCSA, we have selected 12 widely used test problems in the field. 
  They consist of ZDT~\citep{zitzler2000comparison} and DTLZ~\citep{deb2002scalable}.
  Each test suite contains several functional forms and can feature various aspects of optimization algorithms. 
  Comprehensive analysis on the characteristics of the two test suites are well documented by Huband \textit{et al.}~\citep{huband2006review}.
  In both suites, the input vector, $\mathbf{x}$, is divided into two sets $\mathbf{y}$ and $\mathbf{z}$ to construct test problems as follows,
  \begin{align*}
    \text{Given } & \mathbf{x} = \{x_1,\dotsc,x_n\} \\
    \text{let }   & \mathbf{y} = \{y_1,\dotsc,y_j\} = \{x_1,\dotsc,x_j\}\\
    & \mathbf{z}=\{z_1,\dotsc,z_k\} = \{x_{j+1},\dotsc,x_n\}
  \end{align*}
  , where $n$ and $j$ are the dimensions of decision and objective spaces respectively and $k=n-j$.

  \subsection{ZDT}
  {
    The ZDT problem suite consists of six test problems and is probably the most popular test suite to access multiobjective optimization algorithms. 
    The explicit functional forms of five ZDT problems are presented in Table~\ref{tab:zdt}. 
    The ZDT test suite has two main advantages: (a) The Pareto fronts of the problems are known in exact forms and (b) benchmark results of many existing studies are available.
    However, there are shortcomings: (a) The problems have only two objectives, (b) none of the problems contain flat regions and (c) none of the problems have degenerate Pareto optimal front~\citep{huband2006review}.
    
    \begin{table}
      \caption{Five real-valued ZDT problems are described. The first objective depends only on the first decision variable as $f_1(y_1)$ and the second objective is given as $f_2(y_1,\mathbf{z})=g(\mathbf{z})h(f_1(y_1), g(\mathbf{z}))$, where $\mathbf{y}=\{y_1,\dotsc,y_j\} =\{x_1,\dotsc,x_j\}$ and $\mathbf{z}=\{z_1,\dotsc,z_k\} = \{x_{j+1},\dotsc,x_n\}$, where $n$ and $j$ are the dimensions of the decision space and the objective space. Unless the functional forms of $f_1, g$ and $h$ are separately given, they are identical to those of ZDT1.}
      \label{tab:zdt}
      \begin{tabular}{|c|l|c|}
        \hline
        Name & Problem & domains \\
        \hline
        ZDT1 & \usebox{\zdtone} & [0,1] \\
        \hline
        ZDT2 & $h=1-(f_1/g)^2$ & [0,1] \\
        \hline
        ZDT3 & $h=1-\sqrt{f_1/g}-(f_1/g)sin(10\pi f_1)$&[0,1]\\
        \hline
        ZDT4 & $g=1+10k+\sum^{k}_{i=1}(z_{i}^2-10cos(4\pi z_i))$ & \usebox{\zdtfourdomain} \\
        \hline
        ZDT6 & \usebox{\zdtsix}&[0,1]\\
        \hline
      \end{tabular}
    \end{table}
  }
  \subsection{DTLZ}
  {
    The original DTLZ suite consists of nine test problems which are scalable to any number of objectives. 
    This scalability is important since it makes the test suite suitable for testing algorithms for \emph{many} objective problems. 
    The explicit functional forms of the first seven DTLZ problems are presented in Table~\ref{tab:dtlz}.    DTLZ8 and DTLZ9 are omitted in many benchmark studies due to their additional constraints and they are also omitted in this study.

      \begin{table}
        \caption{Seven real-valued DTLZ test problems are described. The objective space is $j$-dimensional and $j$ is set to 3 for benchmarking in this study. The input vectors are $\mathbf{y}=\{y_1,\dotsc,y_j\} = \{x_1, \dotsc, x_j\}$ and $\mathbf{z}=\{z_1,\dotsc,z_k\} = \{x_{j+1},\dotsc,x_n\}$, where $n$ is the dimension of decision space and $k=n-j$.}
        \label{tab:dtlz}
        \begin{tabular}{|c|l|c|}
          \hline
          Name & Problem & domains \\
          \hline
          DTLZ1 & \usebox{\dtlzone} & [0,1] \\
          \hline
          DTLZ2 & \usebox{\dtlztwo} & [0,1] \\
          \hline
          DTLZ3 & Same as DTLZ2, except $g$ is replaced by the one from DTLZ1 & [0,1] \\
          \hline
          DTLZ4 & Same as DTLZ2, except $y_i$ are replaced by $y_i^{0.5}$  & [0,1]\\
          \hline
          DTLZ5 & Same as DTLZ2, except $y_{2,\dotsc,j-1}$ are replaced by $\frac{1+2g y_{i}}{2(1+g)}$. & [0,1] \\
          \hline
          DTLZ6 & Same as DTLZ5, except $g$ is replaced by $g = \sum^{k}_{i=1}z_{i}^{0.1}$.&[0,1] \\
          \hline
          DTLZ7 & \usebox{\dtlzseven} & [0,1] \\
          \hline
        \end{tabular}
      \end{table}
  }
}

\section{Performance measures}
\label{sec:measure}
{
  To evaluate the performance of an algorithm, two aspects are considered: Solutions should be (a) as close to the Pareto front as possible and (b) as diversely distributed among them. 
  Here, we have used four measures defined below.
  \subsection{Spacing ($S$) and average distance between solutions ($\langle d \rangle$)}
  {
    The spacing, $S$, measures how uniformly generated solutions $\mathcal{GS}$ are distributed in the objective space and it is defined as
    \begin{align*}
      S & = \sqrt{\frac{1}{N} \sum^{N}_{i=1} \left( d_{i} - \langle d \rangle\right )^{2}},\\
      \text{where} \quad d_{i} & = \min_{\substack{j \in \mathcal{GS}\\ j \neq i}} \sum^{m}_{k=1}{|f_{k}^{i}-f_{k}^{j}|}, \\
      \langle d \rangle & = \sum^{N}_{i=1}{d_{i}}/N
    \end{align*}
    , where $N$ is the number of solutions in $\mathcal{GS}$, $d_i$ is the distance between the solution $i$ and its nearest neighbor in the objective space in terms of the Hamming distance and $\langle d \rangle$ is the average of $d_i$~\citep{scott1995fault}. Ideally, we want a larger value of $\langle d \rangle$ with a smaller value of $S$.
  }

  \subsection{Generational distance ($GD$)}
  {
    The generational distance measures the average distance between the generated solutions ($\mathcal{GS}$) and the Pareto front~\citep{van2000measuring}.
    $GD$ is defined as 
    \begin{equation}
      GD=\frac{1}{N}\left(\sum^{N}_{i=1}{l^2_i}\right)^{1/2}
    \end{equation}
    , where $N$ is the number of solutions in $\mathcal{GS}$ and $l_i$ is the Euclidean distance between the solution $i \in \mathcal{GS}$ and its nearest in $\mathcal{PF}$. If all solutions in $\mathcal{GS}$ lie on $\mathcal{PF}$, $GD$ is 0. Therefore, a lower $GD$ value is preferred.
  }
  
  \subsection{Error ratio ($ER$)}
  {
    The error ratio measures the fraction of solutions which are not on $\mathcal{PF}$ and it is defined as 
    \begin{equation}
      ER=\frac{\sum^{N}_{i=1}{e_{i}}}{N}
    \end{equation}
    , where $N$ is the number of generated solutions and $e_{i}=0$ if the solution $i$ belongs to $\mathcal{PF}$ within 0.01, $e_i=1$ otherwise~\citep{van2000measuring}.
  }
}

\section{Result}
\label{sec:result}
\begin{figure}[!ht]
\begin{center}
\includegraphics[width=5in]{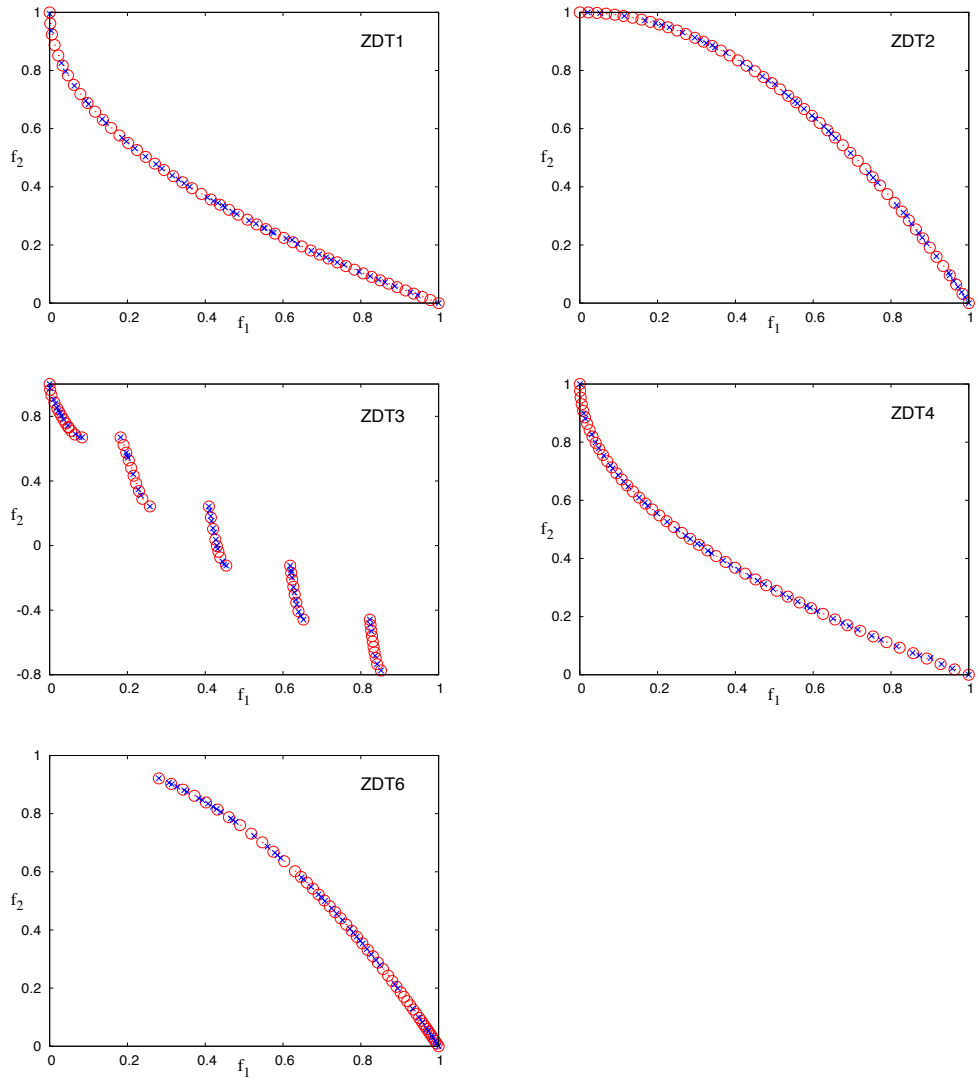}
\end{center}
\caption{
  {\bf Benchmark results on five ZDT problems are shown for MOCSA (red circle) and NSGA2 (blue X)} 
}
\label{fig:ZDT_result}
\end{figure}

\begin{figure}[!ht]
\begin{center}
\includegraphics[width=5in]{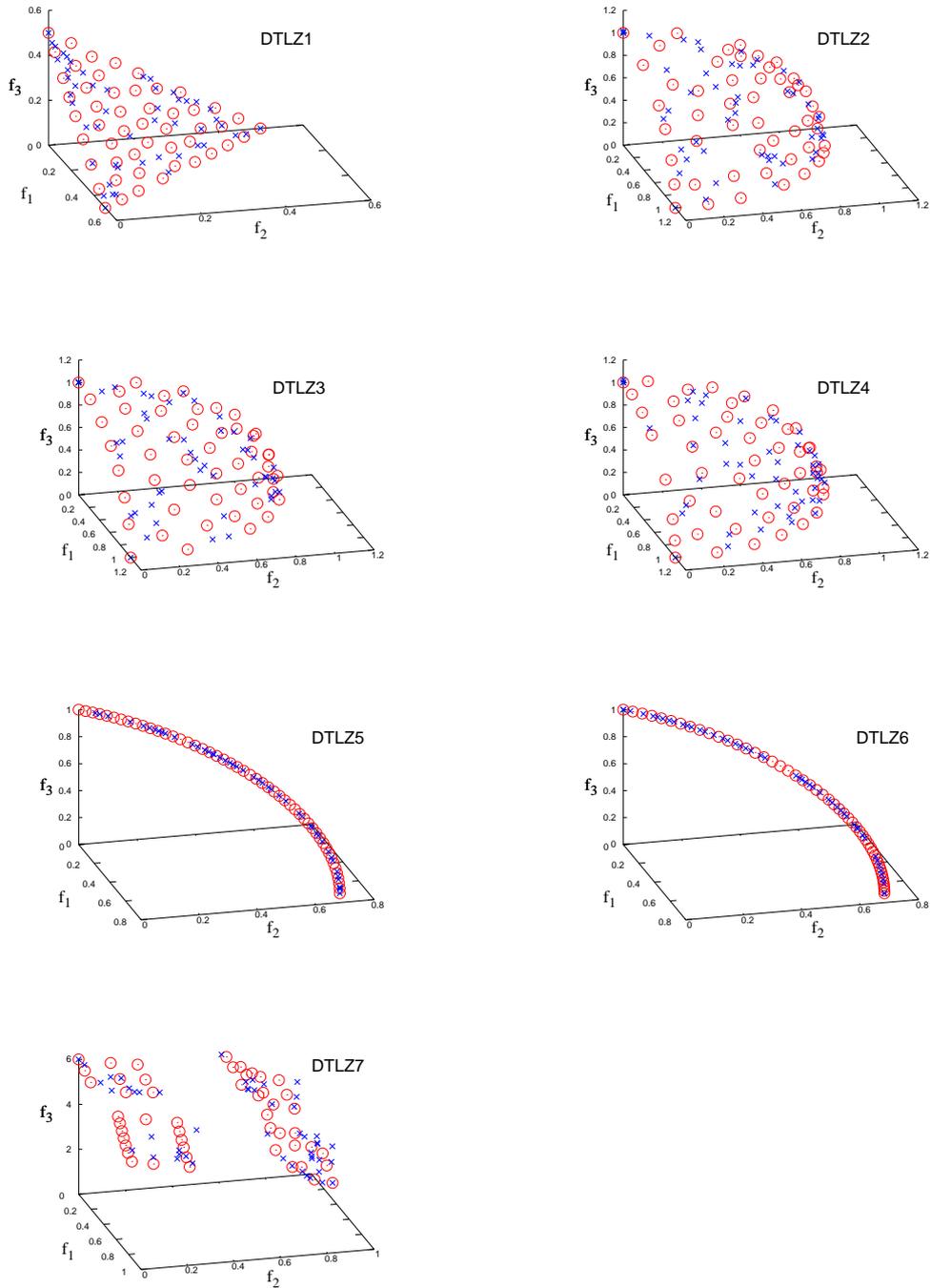}
\end{center}
\caption{
  {\bf Benchmark results on seven DTLZ test problems are shown for MOCSA (red circle) and NSGA2 (blue X)} 
}
\label{fig:DTLZ_result}
\end{figure}

{

  The numerical results on 12 test problems obtained by MOCSA and NSGA2 are shown in Table~\ref{tab:result} and Figures~\ref{fig:ZDT_result} \&~\ref{fig:DTLZ_result}. 
  For benchmarking test, the number of population and trial solutions were set to 50 and 700 for both methods. The NSGA2 was iterated for 250,000 generations and probabilities of crossover and mutation were fixed at 0.9 and 1/24. For MOCSA, the number of seeds used to generate trial solutions was set to 20. The dimensions of decision and objective space of benchmark problems are listed in Table~\ref{tab:problems}. 

  For ZDT problems, MOCSA outperforms NSGA2 in terms of both convergence and diversity of the solutions, except ZDT6. For ZDT6, MOCSA is better than NSGA2 by $\langle d \rangle$ but worse by $S$. For all ZDT problems, entire solutions obtained by MOCSA lies within 0.01 on the Pareto front, shown in red circles in Figure~\ref{fig:ZDT_result}, while NSGA2 solutions are slightly off as shown in Table~\ref{tab:result}. Regarding the spread of the solutions, MOCSA results are more uniformly distributed on the Pareto front, signified by lower S values, than NSGA2 for the first four ZDT problems. 
  
  For DTLZ problems, MOCSA outperforms NSGA2 in almost all aspects. For DTLZ7, 8\% of solutions by MOCSA is not on the Pareto front, while 22\% is the case for NSGA2. In terms of $S$ and $\langle d \rangle$, MOCSA provides more evenly distributed solutions, covering a wider range of the objective space. It should be noted that $\langle d \rangle$ by MOCSA is larger than that by NSGA2 for for all 12 problems tested (by the factor of 1.59 on average), which is signified by the fact that red circles by MOCSA cover Pareto fronts evenly while blue crosses by NSGA2 appear to be locally and unevenly clustered in Figures~\ref{fig:ZDT_result} \&~\ref{fig:DTLZ_result}.
  \begin{table}
    \caption{Dimensions of decision and objective space of benchmark problems are shown.}
    \label{tab:problems}
    \begin{tabular}{|c|c|c|}
      \hline
      Problem & dimension of decision space & dimension of objective space \\
      \hline
      ZDT1-3  & 30 & 2 \\
      ZDT4,6  & 10 & 2 \\
      DTLZ1-7 & 10 & 3 \\
      \hline
    \end{tabular}
  \end{table}
  
  \begin{table}
    \caption{Four performance measures are shown for MOCSA and NSGA2}
    \label{tab:result}
    \begin{tabular}{|c|c|c|c|c|c|c|c|c|c|}
      \hline
      Problem & \multicolumn{4}{c|}{MOCSA} & \multicolumn{4}{c|}{NSGA2} \\
      \cline{2-9}
      & $\langle d \rangle$ & S & GD & ER & $\langle d \rangle$ & S & GD & ER \\ 
      \hline
      ZDT1    & 0.0404 & 0.0055 & 0.0000     & 0     & 0.0270 & 0.0156 & 0.0011    & 0.04      \\
      ZDT2    & 0.0404 & 0.0082 & 0.0000     & 0     & 0.0292 & 0.0146 & 0.0212    & 0.02      \\
      ZDT3    & 0.0438 & 0.0148 & 0.0001     & 0     & 0.0329 & 0.0201 & 0.0020    & 0.02      \\
      ZDT4    & 0.0404 & 0.0097 & 0.0000     & 0     & 0.0328 & 0.0159 & 0.0006    & 0.02      \\
      ZDT6    & 0.0327 & 0.0150 & 0.0000     & 0     & 0.0216 & 0.0119 & 0.0000    & 0         \\
      DTLZ1   & 0.1114 & 0.0068 & 0.0000     & 0     & 0.0615 & 0.0319 & 0.0000    & 0         \\
      DTLZ2   & 0.2319 & 0.0646 & 0.0021     & 0.02     & 0.1361 & 0.0683 & 0.0020    & 0.04      \\
      DTLZ3   & 0.2770 & 0.0225 & 0.0000     & 0     & 0.1139 & 0.0739 & 0.0000    & 0         \\
      DTLZ4   & 0.2478 & 0.0424 & 0.0009     & 0     & 0.1630 & 0.0898 & 0.0019    & 0.02      \\
      DTLZ5   & 0.0487 & 0.0059 & 0.0000     & 0     & 0.0309 & 0.0176 & 0.0610    & 0.06      \\
      DTLZ6   & 0.0484 & 0.0156 & 0.0000     & 0     & 0.0306 & 0.0135 & 0.0000    & 0         \\
      DTLZ7   & 0.2897 & 0.0510 & 0.0011     & 0.04  & 0.1880 & 0.1322 & 0.0071    & 0.22      \\
      \hline
    \end{tabular}
  \end{table}
}

\section{Conclusion}
\label{sec:conclusion}
{
  In this paper, we have introduced a novel multiobjective optimization algorithm by using the conformational space annealing (CSA) algorithm, MOCSA. Benchmark results on 12 test problems show that MOCSA finds better solutions than NSGA2, in terms of four criteria tested. Solutions by MOCSA are closer to the Pareto front, a higher fraction of them are on the Pareto front, they cover a wider objective space, and they are more evenly distributed on average. We note that the efficiency of MOCSA arises from the fact that it controls the diversity of solutions in the decision space as well as in the objective space.
}

\section{Acknowledgement}
\label{sec:acknowledgement}
{
  The authors acknowledge support from Creative Research Initiatives (Center for In Silico Protein Science,
  20110000040) of MEST/KOSEF. We thank Korea Institute for Advanced Study for providing computing
  resources (KIAS Center for Advanced Computation Linux Cluster) for this work. We also would like to
  acknowledge the support from the KISTI Supercomputing Center (KSC-2012-C3-02).
}

\bibliographystyle{elsarticle-harv}







\end{document}